\documentclass[12pt]{article}
\usepackage{cite}
\usepackage[colorlinks]{hyperref}
\usepackage{braket}
\usepackage{mathtools}
\usepackage{mathrsfs}
\usepackage{amsmath}
\usepackage[utf8x]{inputenc}
\usepackage{appendix}
\usepackage{array}
\usepackage{amssymb}
\usepackage{bbm}
\usepackage{bm}
\usepackage{color}
\usepackage[a4paper,top=2cm,bottom=2cm,left=2cm,right=2cm,bindingoffset=5mm]{geometry}
\usepackage{booktabs} 
\usepackage{tikz}
\usepackage{cancel}
\usepackage[bottom]{footmisc}
\usepackage{hyperref}
\usepackage{wrapfig}
\usepackage{placeins}
\usepackage[font=small,skip=-6pt]{caption}

\hypersetup{
colorlinks=true,         
linkcolor=blue,          
citecolor=red,        
urlcolor=red            
}

\newcommand{\ii}{{\rm i}}

\newcommand{\dd}{{\rm d}}

\renewcommand{\vec}[1]{\mathbf{#1}}

\newcolumntype{M}[1]{>{\centering\arraybackslash}m{#1}}

\title{\vspace{-2cm} The Momentum Spaces of $\kappa$-Minkowski noncommutative spacetime}
\date{}
\author{{\Large Fedele Lizzi$^{a,b,c}$\footnote{fedele.lizzi@na.infn.it}, Mattia Manfredonia$^{a,b}$\footnote{mattia.manfredonia91@gmail.com}, Flavio Mercati$^{a,b}$\footnote{flavio.mercati@gmail.com}}
\vspace{12pt}
\\
$^{a}$ Dipartimento di Fisica ``Ettore Pancini'',
\\
Universit\`{a} di Napoli {\sl Federico~II}, Napoli, Italy;
\vspace{6pt}
\\
$^{b}$ INFN, Sezione di Napoli,
\vspace{6pt}
\\
$^{c}$ Departament de F\'{\i}sica Qu\`antica i Astrof\'{\i}sica\\
and Institut de C\'{\i}encies del Cosmos (ICCUB),\\
Universitat de Barcelona, Barcelona, Spain.
}

\begin{document}

\maketitle

\vspace{-12pt}

\begin{abstract}
A useful concept in the development of physical models on the $\kappa$-Minkowski noncommutative spacetime is that of a curved momentum space. This structure is not unique:
several inequivalent momentum space geometries have been identified. Some are associated to a different assumption regarding the signature of spacetime (\emph{i.e.} Lorentzian vs. Euclidean), but there are inequivalent momentum spaces that can be associated to the same signature and even the same group of symmetries. Moreover, in the literature there are two approaches to the definition of these momentum spaces, one based on the right- (or left-)invariant metrics on  the Lie group generated by the $\kappa$-Minkowski algebra. The other is based on the construction of $5$-dimensional matrix representation of the $\kappa$-Minkowski coordinate algebra. Neither approach leads to a unique construction. Here, we find the relation between these two approaches and introduce a unified approach, capable of describing all momentum spaces, and identify the corresponding quantum group of spacetime symmetries. We reproduce known results and get a few new ones. In particular, we describe the three momentum spaces associated to the $\kappa$-Poincar\'e group, which are half of a de Sitter, anti-de Sitter or Minkowski space, and we identify what  distinguishes them. Moreover, we find a new momentum space with the geometry of a light cone, associated to a $\kappa$-deformation of the Carroll group.
\end{abstract}

\newpage

\section{Introduction}

The $\kappa$-Minkowski noncommutative spacetime \cite{Lukierski:1991ff,Lukierski:1992dt,Majid:1994cy} is defined by the commutation relations:
\begin{equation}\label{kappa-Minkowski_timelike}
[x^0,x^i]=\frac{\ii}{\kappa}x^i \,, \qquad \left[x^i,x^j\right]=0 \,, \qquad i, j = 1, \dots ,3,
\end{equation}
where $\kappa$ is a constant with the dimensions of an energy (in $\hbar = 1$ units). The above relations close a Lie algebra, known as $\mathfrak{an}(3)$. The commutation relations~(\ref{kappa-Minkowski_timelike}) can be generalized to $[x^\mu , x^\nu] =\ii ( v^\mu x^\nu -v^\nu x^\mu)$, $\mu = 0, \dots, 3$, where $v^\mu$ is any set of four real numbers. However, all these algebras are isomorphic and can be put in the form~(\ref{kappa-Minkowski_timelike}) by a linear redefinition of generators.  As we will see in Sec.~\ref{SectionEmbeddings} below, this isomorphism does not mean that all algebras with different choices of $v^\mu$ are \emph{physically} equivalent. In fact, the generator $x^0$ is usually interpreted as a time coordinate, and $x^i$ as a spatial one. This interpretation derives from the fact that the above algebra can be derived as the ``quantum homogeneous space'' of a quantum-group deformation of the Poincar\'e group known as $\kappa$-Poincar\'e~\cite{Lukierski:1991ff,Lukierski:1992dt,LUKIERSKI1991331,MajidBicross,Majid:1994cy,Majid:1999td,Lukierski:2015zqa,Lizzi:2019wto,Lizzi:2018qaf} (for a review of fuzzy spacetimes see~\cite{Lizzi:2006bu}, for a different example of quantum homogeneous space see~\cite{Moyalarea}). This group is generated by the elements $a^\mu$ and $\Lambda^\mu{}_\nu$, satisfying the following commutation and cocommutation rules:
\begin{equation}\label{kappaPgroup}
\begin{aligned}
 \Delta[\Lambda^\mu_{\phantom{\mu} \nu}]&=\Lambda^\mu_{\phantom{\mu} \alpha}\otimes\Lambda^\alpha_{\phantom{\alpha} \nu}, &  [\Lambda^\mu_{\phantom{\mu} \nu},\Lambda^\alpha_{\phantom{\alpha} \beta}]&=0,
\\
 \Delta[a^\mu]&=\Lambda^{\mu}_{\phantom{\mu}\nu}\otimes a^\nu+a^\mu\otimes\mathbbm{1}, & [\Lambda ^{\mu}_{\phantom{\mu}\nu},a^\gamma]&= \frac{\ii}{\kappa}\left[\left( \Lambda^\mu_{\phantom{\mu}\alpha}{\delta^\alpha}_0-\delta^\mu{}_0\right)\Lambda^\gamma_{\phantom{\gamma}\nu}+\left( \Lambda^\alpha_{\phantom{\alpha}\nu} \delta^0{}_\alpha -\delta^0{}_\nu\right)\eta^{\mu\gamma}\right],
\\
 S[\Lambda]&=\Lambda^{-1}, ~~ S[a^\mu] = - a^\mu, & [a^0,a^i]&=\frac{\ii}{\kappa}a^i, ~~ [a^i,a^j]=0 ,
\\
 \varepsilon [\Lambda^\mu_{\phantom{\mu}\nu}]&=\delta^\mu{}_\nu  , ~~
\varepsilon [a^\mu]=0 , & 
\Lambda^\mu_{\phantom{\mu}\alpha} \Lambda^\nu_{\phantom{\nu}\beta} \eta^{\alpha\beta} &= \eta^{\mu\nu},  ~~ \Lambda^\rho_{\phantom{\rho}\mu} \Lambda^\sigma_{\phantom{\sigma}\nu} \eta_{\rho\sigma} =\eta_{\mu\nu}, 
\end{aligned}
\end{equation}
where $\eta^{\mu\nu} =\text{diag} ( 1 , - 1 ,- 1 , -1)$ is the usual Minkowski flat metric. Note that the commutation and cocommutation rules involving the Lorentz sector are undeformed, the deformation being limited to the translation sector, and the mixed part. The very last relation gives the appropriate number of constraints so that the independent components of $\Lambda$ are six.
The $\kappa$-Minkowski commutation relations~(\ref{kappa-Minkowski_timelike}) are left invariant by the following left co-action of $\kappa$-Poincar\'e :	
\begin{equation}
\Delta_L[x^\mu]=\Lambda^\mu_{\phantom{\mu} \nu}\otimes x^\nu+a^\mu\otimes\mathbbm{1},
\end{equation}
which is an algebra homomorphism for the relations~(\ref{kappa-Minkowski_timelike}). This is the sense in which $\kappa$-Minkowski is the quantum homogeneous space associated to $\kappa$-Poincar\'e, and, in this light, it is legitimate to interpret $x^0$ as temporal and $x^i$ as spatial coordinates, because they transform as such, and the separation between time and space indices in the generators of the  $\kappa$-Poincar\'e group is determined by the form of the metric $\eta^{\mu\nu}$ appearing in its relations.  In Sec.~\ref{SectionEmbeddings} we will show that any algebra $[x^\mu , x^\nu] =\ii ( v^\mu x^\nu -v^\nu x^\mu)$ is covariant under a generalization of the quantum group~(\ref{kappaPgroup}), with any choice of the matrix $\eta_{\mu\nu}$. A linear redefinition of generators $x^\mu$,  done to show that all algebras with different values of $v^\mu$ are isomorphic, induces a linear transformation of the matrix $\eta_{\mu\nu}$. Hence, to specify a model of quantum spacetime, the commutation relations of the coordinates $x^\mu$ (\emph{i.e.} the choice of parameters $v^\mu$) are not sufficient: one needs also to specify the form of the matrix $\eta_{\mu\nu}$, which can attribute timelike or lightlike nature to different linear combinations of generators $x^\mu$. Two models, specified by $(v^\mu , \eta_{\mu\nu})$  and $(v'^\mu , \eta'_{\mu\nu})$  will then be physically equivalent if there exists a linear transformation of the generators that sends $v^\mu$ to $v'^\mu$  and, simultaneously, $\eta_{\mu\nu}$  to $\eta'_{\mu\nu}$.

One unusual property of the $\kappa$-Minkowski spacetime is that it is associated to a \emph{curved momentum space}, a momentum space that generalizes the vector space of Special/General Relativity into a pseudo-Riemannian geometry.
 The idea of a curved momentum space has a long history. It was formulated more than 20 years ago by S. Majid (see, \emph{e.g.},  \cite{Majid:1999td,Majid:2006xn,Majid:1988we}), and it can be mathematically understood as dual in the sense of Hopf-algebraic duality of algebraic and coalgebraic sectors describing  quantum coordinates and  quantum momenta; the noncommutativity of translation generators is a manifestation of the curvature of spacetime. Similarly, in a quantum geometry, the noncommutativity of spacetime coordinates is a manifestation of the curvature of momentum space, a phenomenon that Majid called `cogravity'.

The geometrical structure of the $\kappa$-Minkowski momentum space was first studied in~\cite{KowalskiGlikman:2003we}, and then further studied in a variety of works, including~\cite{KowalskiGlikman:2002ft,KowalskiGlikman:2003we,Freidel:2007hk,Arzano:2009ci,Arzano:2010kz,Arzano:2014jfa,Gutierrez-Sagredo:2019ipf}. The simplest way to see this is to consider the \emph{ordered plane waves} built from the noncommutative coordinates~(\ref{kappa-Minkowski_timelike}):
\begin{equation}\label{WeylOrderedExponentials}
\exp \left(\ii k_\mu x^\mu \right) \,, \qquad k_\mu \in \mathbbm{R}^4\,.
\end{equation}
these are useful because they provide a basis in which we can expand functions, in order to discuss field theories on $\kappa$-Minkowski~(\ref{kappa-Minkowski_timelike})~\cite{Kosinski2000,Agostini:2002yd,Agostini:2002de,Agostini:2005mf,Arzano:2017opw,Mercati:2011aa,Mercati:2018hlc,Mercati:2018ruw}. What is unusual is the fact that, because the product~(\ref{kappa-Minkowski_timelike}) is noncommutative, these plane waves do not combine in a linear way:
\begin{equation}
\begin{aligned}
   & \exp\left( \ii k_\mu x^\mu \right)\exp\left( \ii q_\mu x^\mu \right)  = \\ 
   &\exp \left\{  \ii  \frac{(k_0+q_0)/\kappa} {e^{(k_0+q_0)/\kappa} -1}\left[ \left( \frac{e^{k_0/\kappa} -1}{k_0/\kappa}\right)  k_i + e^{-k_0/\kappa}\left( \frac{e^{q_0/\kappa} -1}{q_0/\kappa}\right)  q_i\right]  x^i + \ii (k_0+q_0) x^0 \right\}) .
    \end{aligned}
\end{equation}
%
This fact,  known since the 1990s~\cite{AmelinoCamelia:1999pm}, can be proven explicitly using only the commutation relations~\cite{Agostini:2005mf,Mercati:2011aa}.  Since the commutation rules are those of a Lie algebra, the exponentials are closed under product, they form a subalgebra of the universal enveloping algebra of $\mathfrak{an}(3)$. The law:
\begin{equation}
     (k,q) \to p  \,, ~~ p_i = \frac{(k_0+q_0)/\kappa} {e^{(k_0+q_0)/\kappa} -1}\left[ \left( \frac{e^{k_0/\kappa} -1}{k_0/\kappa}\right)  k_i + e^{-k_0/\kappa}\left( \frac{e^{q_0/\kappa} -1}{q_0/\kappa}\right)  q_i\right]\,, ~ p_0 = k_0+q_0 \,,
\end{equation}
generalizes in a nonlinear way the familiar composition law of ``wave vectors'' (or Fourier parameters) $(k,q) \to k_\mu + q_\mu$, and reduces to it in the limit $\kappa \to \infty$. It can be seen as a small deformation of it, when the wave vectors are much smaller than $\kappa$~\cite{Carmona:2011wc,Carmona:2012un}. There is a consensus in the literature on the fact that this nonlinearity is a manifestation of the fact that the Fourier parameters are coordinates on a nonlinear manifold. In fact, exponentiating the generators of a Lie algebra like $\mathfrak{an}(3)$, one obtains elements of the associated Lie group, which in our case is $AN(3)$~\cite{Arzano:2010kz}. Then, since the algebra is not Abelian, the composition law between the parameters  in the exponentials is not linear, and they just codify the group product. As the theory of Lie groups prescribes, these parameters are coordinate systems on the group manifold.

The expression of plane waves in~(\ref{WeylOrderedExponentials}) is not the only possible choice to represent a plane wave, we had implicitly chosen an ordering prescription. There are other ordering choices, which give rise to different factorizations of the group elements. For example, the time generator can be ordered to the right, $\exp(\ii q_i x^i) \exp( \ii q_0 x^0)$. Different orderings are related through nonlinear relations between the real parameters appearing in the exponentials. For the two examples above:
\begin{equation}
    \exp \left( \ii k_\mu x^\mu \right) = 
        \exp \left[ \ii \left( \frac{e^{k_0 / \kappa} -1 }{k_0 / \kappa} \right) k_i  x^i \right]  \exp \left( \ii k_0 x^0 \right) \, ,
\end{equation}
this transformation, $k_0 \to k_0$, $k_i \to \left( \frac{e^{k_0 / \kappa} -1 }{k_0 / \kappa} \right) k_i $ is a general coordinate change, \emph{i.e.}, a diffeomorphism on the group manifold.

We interpret the group manifold associated to the Lie group $AN(3)$ as the momentum space of theories on $\kappa$-Minkowski that make use of noncommutative plane waves, \emph{e.g.} (quantum) field theories, in which ordered plane waves are a basis for scalar fields and solutions of the equations of motion. Here we are interested in the \emph{geometry} of this momentum space. In Lie group theory, there is a natural way to define a metric on the group manifold: if there is a nondegenerate Killing form, one can immediately define a bi-invariant metric. However, since the group $AN(3)$ is not semi-simple, the Killing form is degenerate and there is no bi-invariant metric. There is, however, a basis of left-invariant forms and another one of right-invariant forms:
\begin{equation}
\begin{gathered}
\theta_0^L = d k_0 \,, \qquad  \theta_i^L = d k_i  + \frac{k_i}{\kappa} d k_0\,,
\\
\theta_0^R = d k_0 \,, \qquad  \theta_i^R = e^{k_0/\kappa} d k_i \,,
\end{gathered}
\end{equation}
which are related by a diffeomorphism implementing the group inverse (codified by the antipode map at the Hopf algebra level): $k_0 \to - k_0$, $k_i \to - e^{k_0/\kappa} k_i$, which sends $\theta_\mu^L \to  -\theta_\mu^R$ and vice-versa. As observed in~\cite{DanijelJurman2017}, any quadratic form built from the symmetrized tensor product of right-invariant forms will give a right-invariant metric, and the same for left-invariant metrics.
In~\cite{DanijelJurman2017} it is stated that there is a unique right-invariant metric and a unique left-invariant one, but the author is implicitly assuming that the signature is $(+,-,-,-)$, and the rank is maximal (no zero eigenvalues). Moreover, even under these conditions, there is a hidden assumption in~\cite{DanijelJurman2017}'s proof that all metrics are diffeomorphic. In the following Section we will study in detail all the possible right-invariant metrics, and classify them according to diffeomorphism classes.

In the literature on $\kappa$-Minkowski, the most common way used to introduce the curved momentum space is that used in~\cite{KowalskiGlikman:2003we}, which was the first paper to observe that the momentum space of $\kappa$-Minkowski may be described as a curved Riemannian manifold. \cite{KowalskiGlikman:2003we} used a matrix representation of $\mathfrak{an}(3)$ (see also~\cite{Ballesteros:2017pdw,Ballesteros:2019bnc,Ballesteros:2019hbw}). The algebra~(\ref{kappa-Minkowski_timelike}), in fact, can be seen as a subalgebra of the five-dimensional Lorentz algebra $\mathfrak{so}(4,1)$ via the isomorphism:
\begin{equation}\label{SO(4,1)_algebra_embedding}
    x^\mu \sim M^{0\mu} + M^{4\mu} \,,
\end{equation}
where $M^{AB}\, (A,B=0,\ldots,4)$ are the Lorentz generators in the standard antisymmetric $5 \times 5 $ matrix representation. 
This isomorphism induces the following five-dimensional representation of the commutation relations~(\ref{kappa-Minkowski_timelike}):
\begin{equation}\label{Representation_of_kappa_Minkowski}
\rho(x^0) = - \frac \ii \kappa \left( 
\begin{array}{ccc}
0 & \vec{0} & 1
\\
\vec{0} & \hat{0} & \vec{0}
\\
1 & \vec{0} & 0
\end{array}
\right) \,, \qquad 
\rho(x^i) = - \frac \ii \kappa \left( 
\begin{array}{ccc}
0 & \vec{e}_i & 0
\\
\vec{e}_i & \hat{0} & \vec{e}_i
\\
0 & -\vec{e}_i & 0
\end{array}
\right) \,,
\end{equation}
where $\vec{e}_i^a =\delta^a_i$, three-dimensional vector quantities are in boldface (we do not distinguish between rows and colums, it should be clear form the position in the matrix), $\hat{0}$ is the zero $3\times 3$ matrix. This is a $*$-representation under the involution compatible with the Lorentz group $(\rho^\alpha{}_\beta)^* = \eta^{\alpha\lambda} \eta_{\gamma\beta} \overline{\rho^\gamma{}_\lambda}$ (\emph{i.e.} rising an index, flipping indices, complex conjugating and lowering back the index), which leaves all generators $\rho(x^\mu)$ invariant.
In this way, the plane waves/group elements are represented as the matrices (in order to get simpler formulas we use the ``time-to-the-right'' ordering):
\begin{equation}\label{Exponentialization_k-representation}
G^*(p_\mu) = e^{ \ii p_i \rho(x^i)}e^{ \ii p_0 \rho(x^0)} = 
\left( 
\begin{array}{ccc}
\cosh \frac{p_0}{\kappa} +  e^{\frac{p_0}{\kappa}} \frac{\| \vec p \|^2}{2\kappa^2}
&
\frac{\vec{p}}{\kappa}
&
\sinh \frac{p_0}{\kappa} +  e^{\frac{p_0}{\kappa}} \frac{\| \vec p \|^2}{2\kappa^2}
\\ \\
 e^{\frac{p_0}{\kappa}} \frac{\vec{p}}{\kappa}
 &
\mathbbm{1}
 &
  e^{\frac{p_0}{\kappa}} \frac{\vec{p}}{\kappa}
\\ \\
 \sinh \frac{p_0}{\kappa} -  e^{\frac{p_0}{\kappa}} \frac{\| \vec p \|^2}{2\kappa^2}
&
-\frac{\vec{p}}{\kappa}
&
\cosh \frac{p_0}{\kappa} -  e^{\frac{p_0}{\kappa}} \frac{\| \vec p \|^2}{2\kappa^2}
\end{array}
\right) \,.
\end{equation}
The idea is that, since the above representation is free, and the nondegenerate orbits have the dimension of the group, these will be diffeomorphic to the group manifold.\footnote{For example, exponentiating the standard representation of $\mathfrak{su}(2)$ as $2\times2$ complex matrices acting on the vector space of 2D spinors $\mathbb{C}^2$, one can prove that the nondegenerate orbits of the group are all 3-spheres embedded in $\mathbb{R}^4$ (under the canonical identification $\mathbb{R}^4\sim\mathbb{C}^2$), and indeed the group manifold of $\mathfrak{su}(2)$ is, topologically, a 3-sphere.}
The group orbits can be obtained by taking a fiducial vector $u^A$ in the five-dimensional vector space on which~(\ref{Exponentialization_k-representation}) acts, and considering the points obtained by acting upon $u$ with $G^*(p_\mu)$ for all choices of $p^\mu$:
\begin{equation} \label{build}
X^A = X^A(p_\mu) = G^*(p_\mu)^A{}_B u^B \,.
\end{equation}
$X^A(p_\mu)$ are the parametric representation of a four-dimensional submanifold embedded in a five-dimensional Minkowski space. This submanifold is diffeomorphic to the group manifold of $AN(3)$, and to momentum space.
Since all $G^*(p_\mu)$ are elements of $SO(4,1)$, we have that
\begin{equation}\label{EmbeddingEquationSO(4,1)}
X^A X_A = X^A(p) X^B(p) \eta_{AB} = u^A u^B \eta_{AB}, \qquad    \eta_{AB}= \text{diag}(1,-1,-1,-1,-1) , 
\end{equation}
for all $p^\mu \in \mathbb{R}^4$. Choosing $u^A= (0,0,0,0,1)$, the above equation is that of de Sitter spacetime. The conclusion in~\cite{KowalskiGlikman:2003we} or~\cite{Ballesteros:2017pdw} is that the geometry of momentum space is de Sitter. And indeed the metric on the orbit induced by the embedding $X^A(p)$:
\begin{equation}\label{InducedMetric}
\dd s^2=-\frac{\partial X^A}{\partial p_\mu}\frac{\partial X^B}{\partial p_\nu}\eta_{AB}   \dd p_{\mu} \dd p_{\nu} = \frac 1 {\kappa^2} \left( - \dd p_0^2 + e^{2 p_0 /\kappa}\sum_{i=1}^3 \dd p_i^2 \right),
\end{equation}
is the same right-invariant metric found by~\cite{DanijelJurman2017}. Moreover, one can check that  for $u^A= (0,0,0,0,1)$ the relation $X^0 + X^4 >0$ is verified for all choices of $p_\mu$, and therefore we are actually dealing with half of de Sitter spacetime, the half covered by the flat slicing (the coordinates $p_\mu$ corresponding to time-to-the-right ordering of plane waves are what cosmologists call comoving coordinates for de Sitter spacetime). This constraint makes the portion of momentum space covered by the $p_\mu$ coordinates non-Lorentz-invariant~\cite{Freidel:2007hk}, and one has to choose a different global topology for the ambient space (the \emph{elliptic} topology~\cite{Verlinde}) in order to restore Lorentz invariance~\cite{Mercati:2018ruw,Mercati:2018fba}.

The construction we just described, however, is not unique. For example, in~\cite{Arzano:2014jfa} it was noticed that a different fiducial vector [in particular a time-like one $u^A=(1,0,0,0,0)$] gives rise to a different momentum space. 
Since the Lorentz group has disconnected orbits, corresponding to different fiducial vectors, their choice is not inconsequential. 
Moreover, as it turns out, the representation~(\ref{SO(4,1)_algebra_embedding}) is not the only possible for $\mathfrak{an}(3)$. In~\cite{Blaut2004}, the authors notice that $x^\mu$ can be represented as the four generators ${\hat M}^{4\mu}$ (times $\kappa$) of a five-dimensional orthogonal algebra 
\begin{equation}\label{5DOrthogonalAlgebra}
  [{\hat M}^{AB},{\hat M}^{CD}] = \ii \left ({\hat g}^{BC}{\hat
M}^{AD} - {\hat g}^{AC}{\hat M}^{BD} - {\hat g}^{BD}{\hat M}^{AC}
+ {\hat g}^{AD}{\hat M}^{BC} \right ) \,,
\end{equation}
which preserves a 5D metric of the form $\hat g^{4A} = \hat g^{A4} = \delta^A{}_0$, and $\hat g^{\mu\nu}$ is assumed to have eigenvalues $(-1,1,1,1)$. Then, according to the form of $\hat g^{\mu\nu}$, one has different situations. The 5D matrix can only have the following signatures: $(-,+,+,+,+)$, in which case $\hat M^{AB}$ close the $\mathfrak{so}(4,1)$ algebra, or $(-,-,+,+,+)$, corresponding to the $\mathfrak{so}(3,2)$ algebra, or, finally, $(-,+,+,+,0)$, which implies that $\hat M^{AB}$ generate the Poincar\'e algebra $\mathfrak{iso}(3,1)$. The form of $\hat g^{AB}$ is such that the transformations generated by $\hat M^{\mu\nu}$ (which close a 4D Lorentz algebra) leave the direction $(0,0,0,0,1)$ in the ambient 5D space invariant, while those generated by $\hat M^{4\mu}$ change it. Therefore, in~\cite{Blaut2004}, the momentum space is identified with the quotient of the 5D group with $SO(3,1)$, which, according to the choice of matrix  $\hat g^{\mu\nu}$ will be de Sitter [$SO(4,1)/SO(3,1)$], anti-de Sitter de Sitter [$SO(3,2)/SO(3,1)$] or Minkowski [$ISO(3,1)/SO(3,1)$].

However, this method allows us only to know the local isometries of momentum space, and does not reveal its global shape. For example, in the de Sitter case, it is necessary to explicitly study the submanifold of $\mathbbm{R}^5$ that is traced by the action of the generators $x^\mu$, as was illustrated in Eqs.~(\ref{SO(4,1)_algebra_embedding}-\ref{InducedMetric}), in order to realize that it is only half of the de Sitter hyperboloid that can be identified with momentum space.

The purpose of the present  paper is to  study all the momentum spaces that can be associated to $\kappa$-Minkowski, and to find the relation between the different partial analyses of the past, \emph{e.g.} \cite{KowalskiGlikman:2003we,Blaut2004,Arzano:2014jfa,DanijelJurman2017}.
A synopsis of the main result is:
\begin{itemize}
    \item In Sec.~\ref{Right-invariant-metrics-sec} we study all the possible right-invariant metrics on the $AN(3)$ group, and their equivalence classes under diffeomorphisms, which has not been done before.
    
    \item In Sec.~\ref{SectionEmbeddings} we clarify the sense in which different $\kappa$-Minkowski algebras (with different choices of parameters $v^\mu$ in the commutation relations) can be said to be physically equivalent or inequivalent.
    
    \item In Sec.~\ref{SectionMomentumSpaceGeometries} we study all \emph{nondegenerate} geometries of the $\kappa$-Momentum spaces, which can be obtained with different choices of $v^\mu$ parameters \emph{and} spacetime metric. Many of these were already known in the literature (de Sitter, anti-de Sitter, Riemannian hyperbolic). We find two new ones:
    
    \begin{itemize}
        \item  a hyperbolic space with signature $(+,+,-,-)$, associated to a $\kappa$-Minkowski spacetime symmetric under a $\kappa$-deformation of the $ISO(2,2)$ group, 
        
        \item a cone, associated to a $\kappa$-Minkowski spacetime symmetric under a $\kappa$-deformation of the Carroll group (or of a Lorentzian version of the Carroll group),
        
        \item half of Minkowski space.
    \end{itemize}
    
Furthermore, we complete the classification of the momentum spaces associated to $\kappa$-Minkowski spacetimes symmetric under the $\kappa$-Poincar\'e group: these can be (half of) de Sitter, (half of) Minkowski or (half of) anti-de Sitter, depending on which coordinates the spacetime metric $g^{\mu\nu}$ makes timelike: in the first case $x^0$, in the second $x^0 \pm x^i$ for some $i$, and in the third $x^i$ for some $i$.

\end{itemize}

\section{Right-invariant metrics on $AN(3)$  \label{Right-invariant-metrics-sec}
}

  There is a certain freedom in choosing these metrics, and we need to classify them and find the physical meaning of inequivalent classes.

 Let us analyze the most generic right-invariant metric similarly to what was done in~\cite{DanijelJurman2017}, but without any initial assumption. One begins with the generic right-invariant line element:
\begin{equation} \label{General_R-invariant_metric}
ds^2 = -g^{\mu\nu} \theta_\mu^R \theta_\nu^R \,,
\end{equation}
where $ g^{\mu\nu}$ is a generic symmetric $4 \times 4 $ matrix. Calculating the Riemann and Ricci curvatures, one can verify that, whatever $ g^{\mu\nu}$  we choose, the curvatures satisfy the relation of maximally-symmetric spaces:
\begin{equation}
R_{\mu\nu\rho\sigma} = \frac 1{12} R \left( g_{\mu \rho} g_{\nu \sigma} - g_{\mu\sigma} g_{\nu\rho} \right) \,,
\end{equation}
and the scalar curvature is
\begin{equation} \label{RicciScalar}
R = \frac{12 \det ( g^{ij}) }{\det ( g^{\mu\nu} )} \,.
\end{equation}
This result allows us to connect the approach based on right-invariant metrics
to that of~\cite{Blaut2004} (based on a family of $5D$ representations of $\mathfrak{an}(3)$): both approaches agree that the local geometry of the momentum spaces of $\kappa$-Minkowski is that of a maximally-symmetric space. Clearly, if the momentum-space metric is assumed to be non-singular, the only possibilities are de Sitter, anti-de Sitter and Minkowski space.

Let us study in more detail what classes of inequivalent metrics (under coordinate changes) we have, without assuming anything on the signature or singularity of the 4D metric.  We begin by  decomposing the indices in Eq.~(\ref{General_R-invariant_metric}) in $0$ and $i=(1,2,3)$:
\begin{equation}
ds^2 = - g^{00} d k_0^2  - 2 e^{k_0/\kappa} g^{0i} dk_0 dk_i  - e^{2 k_0/\kappa} g^{ij} dk_i dk_j \,,
\end{equation}
we can introduce logarithmic coordinates as $k_0 = - \kappa \log \tau $ ($k_0 \in \mathbbm{R}$ so $\tau \in \mathbbm{R}^+$):
\begin{equation}\label{MetricInLogCoords}
ds^2 =  - \frac 1 {\tau^2}\left( g^{00} d \tau^2  - 2 g^{0i} d \tau  dk_i  +  g^{ij} dk_i dk_j \right) \,,
\end{equation}
if $g^{0i}=0$, the metric is already block-diagonal, and a simple linear transformation of the $k_i$,  $i=1,2,3$ coordinates diagonalizes $g^{ij}$ and puts the line element in the form~(\ref{GeneralRightInvariantMetric}). If any of the three components $g^{0i}$ are nonzero, we can perform the following linear transformation: $k_i = k'_i + c_i \, \tau$, and the line element takes the form
 \begin{equation}
ds^2 = - \frac 1 {\tau^2}\left[ ( g^{00} - 2 g^{0i} c_i + g^{ij} c_i c_j) d \tau^2 + 2(c_i g^{ij}  - g^{0j} ) dk'_j d\tau + g^{ij} dk'_i dk'_j \right] \,,
\end{equation}
and the line element can be block-diagonalized if the following equations can be solved for  $c_i$:
\begin{equation}\label{EqFor_c}
g^{ij} c_j =  g^{0i} \,.
\end{equation}
The above equation admits solutions only if the range of $g^{ij}$ is equal to the range of the rectangular matrix $g^{\mu j}$, which means that the 3-vector $g^{0i}$ lies within the range of $g^{ij}$. This is always true if $g^{ij}$ has maximum rank (\emph{i.e.} it is invertible), but if $g^{ij}$ has some zero eigenvalues it might not hold. Assuming that $g^{0i}$ lies within the range of $g^{ij}$, we can replace the solution of (\ref{EqFor_c})  into the line element, using the fact that $g^{0i} c_j = g^{ij} c_i c_j$ :
 \begin{equation}
ds^2 = - \frac 1 {\tau^2}\left[ ( g^{00} - g^{ij} c_i c_j) d \tau^2 + g^{ij} dk'_i dk'_j \right] \,,
\end{equation}
and, finally, we can perform a linear transformation of the $k'_i$ coordinates that diagonalizes the matrix $g^{ij}$:
 \begin{equation}
ds^2 = - \frac 1 {\tau^2}\left[ \lambda^0  d \tau^2 + \left( \lambda^1  (dk''_1)^2  + \lambda^2  (dk''_2)^2 \right)  +\lambda^3  (dk''_3)^2   \right] \,,
\end{equation}
where $\lambda^0  =g^{00} - \lambda^1 (c'_1)^2  - \lambda^2 (c'_2)^2 -\lambda^3 (c'_3)^2$, and $c'_i$ is the transformed version of $c_i$. Going back to the $k_0$ variable:
 \begin{equation} \label{GeneralRightInvariantMetric}
ds^2 =-  \lambda^0 \, d k_0^2  - e^{2k_0/\kappa} \left[ \lambda^1 (dk''_1)^2 + \lambda^2 (dk''_2)^2  + \lambda^3 (dk''_3)^2 \right] \,,
\end{equation}
where the coefficients $\lambda^\mu$ are unconstrained real numbers.

Assume now that Eq.~(\ref{EqFor_c}) is not solvable: then $g^{ij}$ has some zero eigenvalues, and we know that the geometry is flat, because the Ricci scalar is proportional to $\det (g^{ij}) / \det (g^{\mu\nu})$  [Eq.~(\ref{RicciScalar})], and sending one, two or three of the eigenvalues of $g^{ij}$ simultaneously to zero makes $R$ vanish (no matter what the eigenvalues of $g^{\mu\nu}$ do). Moreover, the fact that Eq.~(\ref{EqFor_c}) is not solvable implies that  $g^{0i}$ has nonzero components in the zero eigenspace of $g^{ij}$. We need to consider two distinct cases:
\begin{itemize}

\item If $g^{00}=0$, then the metric~(\ref{MetricInLogCoords}) takes the form
\begin{equation}
ds^2 =  - \frac 1 {\tau^2} \left[  2  g^{0i} d\tau dk_i  + g^{ij} dk_i dk_j \right] \,,
\end{equation}
and transforming the spatial indices in order to diagonalize the spatial part:
\begin{equation}
ds^2 =  - \frac 1 {\tau^2} \left[  2 v^i d \tau dk'_i  + \left( \lambda^1  (dk'_1)^2  + \lambda^2  (dk'_2)^2  +\lambda^3  (dk'_3)^2 \right) \right] \,,
\end{equation}
where some of the diagonal elements $\lambda^i$ are zero (and the 3-vector $v^i = g'^{0i}$ has at least one nonzero component in the direction corresponding to a zero eigenvalue). The determinant of this metric is:
\begin{equation}
- \tau^{-8} \left( \lambda^2\lambda ^3 (v^1)^2 + \lambda ^1\lambda ^2 (v^3)^2 + \lambda ^1\lambda ^3 (v^2)^2 \right) \,,
\end{equation}
which is zero unless only one eigenvalue is zero. If, for example, $\lambda^1=0$ (and consequently $v^1 \neq 0$), and $\lambda^2 , \lambda^3 \neq 0$, then the determinant is $- \tau^{-8} v^1 \lambda^2 \lambda^3 \neq 0$, and the 4D metric has four nonzero eigenvalues.

If there are two zero $\lambda$'s, for example $\lambda^1$ and $\lambda^2$, then the characteristic polynomial of the 4D metric reduces to $q^4  - \lambda_3 q^3 - \| \vec v \|^2 q^2 + \lambda^3 [ (v^1)^2 + (v^2)^2] q $, and since $v^1,v^2,\lambda^3 \neq 0$, there are three nonzero eigenvalues.

Finally, if all three $\lambda^i$ are zero, the 4D metric has two zero eigenvalues, and the remaining two are $\pm \| \vec v\|$.


\item If $g^{00} \neq 0$, we can make the coordinate transformation $\tau = \theta + g^{0i}/g^{00}  k_i$,
\begin{equation}
ds^2 =  - \frac 1 {(\theta + g^{0i} k_i / g^{00})^2} \left( g^{00} d \theta^2  +   (g^{0i}g^{0j}/g^{00} + g^{ij}) dk_i dk_j \right) \,.
\end{equation}
The matrix $\left( g^{0i}g^{0j}/g^{00} + g^{ij} \right)$ has determinant
\begin{equation}
\lambda^2 \lambda^3 (v^1)^2 +
\lambda^1 \lambda^2 (v^2)^2+
\lambda^1 \lambda^3 (v^3)^2 - 
\lambda^1 \lambda^2 \lambda^3
\end{equation}
(where $\lambda^i$ are the eigenvalues of $g^{ij}$). This, again, can be nonzero if only one of the  $\lambda^i$ is zero. 
In the other two cases, just like before, one has one or two zero eigenvalues.
\end{itemize}
We conclude that, when Eq.~(\ref{EqFor_c}) is not solvable, one has a flat 4D metric of arbitrary signature with two, three or four nonzero eigenvalues.

Regarding the left-invariant metric, by what we said before they are all diffeomorphic to the right-invariant ones, so they just correspond to the same geometries described here, written in different coordinates.

\section{Embeddings in 5D-metric-preserving groups}\label{SectionEmbeddings}

$\hat M^{AB}$ closing the algebra~(\ref{5DOrthogonalAlgebra}), which preserves a 5D metric of the form: 
\begin{equation}\label{5DMatrix_hat_g}
\hat{g}^{AB} = 
\left(
\begin{array}{ccccc}
g^{00} & \dots &&g^{03}& 1 
\\
\vdots &  \ddots && \vdots & 0
\\
&&&& 0
\\
g^{03} & \dots &&g^{33}& 0
\\
1 & 0 & 0 & 0 & 0
\end{array}
\right) \,,
\end{equation}
introducing the representation $\rho(x^\mu) = \frac 1 \kappa M^{4 \mu}$ one can immediately verify that the commutation relations~(\ref{5DOrthogonalAlgebra}) imply the $\kappa$-Minkowski commutation relations, $[ \rho(x^\mu) , \rho(x^\nu) ] = \frac \ii \kappa \left( \delta^\mu{_0}   \rho(x^\nu) - \delta^\nu{_0} \rho( x^\mu) \right )$, thanks to the particular form of $\hat g^{AB}$. Similarly to what we illustrated in the introduction 
for the representation~(\ref{Representation_of_kappa_Minkowski}), we can now calculate the generic group elements generated by $\rho(x^\mu)$, whic are a four-parameter family of 5D matrices $G^*(p_\mu) = e^{ \ii p_i \rho(x^i)}e^{ \ii p_0 \rho(x^0)}$.

Choosing a fiducial vector $u^A$ in the ambient space, we then derive its orbits as $X^A = X^A(p_\mu) = G^*(p_\mu)^A{}_B u^B$. Now, assuming that $\hat g^{AB}$ is invertible and calling its inverse $\hat g_{AB}$, we can find the metric that is induced on the orbit by the embedding $X^A(p_\mu)$ as
\begin{equation}
\dd s^2=- \hat g_{AB} \frac{\partial X^A}{\partial p_\mu}\frac{\partial X^B}{\partial p_\nu}  \dd p_{\mu} \dd p_{\nu} \,.
\end{equation}
From the definition of right-invariant forms, we know that
\begin{equation}
({G^*}^{-1})^A{}_B \, d {G^*}^B{}_C = \theta^R_\mu \rho(x^\mu)^A{}_C \,,
\end{equation}
which implies that 
\begin{equation}
\frac{\partial X^A}{\partial p_\mu} dp_\mu = d {G^*}^A{}_B u^B
= \theta^R_\mu ({G^*}^{-1})^A{}_C  \rho(x^\mu)^C{}_B u^B \,.
\end{equation}
Now, assuming that $\hat g^{AB}$ is invertible, by definition its inverse is left invariant by the group elements:
\begin{equation}
\hat g_{AB} ({G^*}^{-1})^A{}_C ({G^*}^{-1})^B{}_E=  \hat g_{CE} \,,
\end{equation}
and so we can write 
\begin{equation}
\begin{aligned}
\dd s^2 &=- \hat g_{AB} \frac{\partial X^A}{\partial p_\mu}\frac{\partial X^B}{\partial p_\nu}  \dd p_{\mu} \dd p_{\nu} = - 
\theta^R_\mu \theta^R_\nu \, \hat g_{AB} ({G^*}^{-1})^A{}_C ({G^*}^{-1})^B{}_E  
\rho(x^\mu)^C{}_D 
 \rho(x^\nu)^E{}_F u^Du^F
\\
&= - 
\theta^R_\mu \theta^R_\nu \,
\hat g_{CE} 
\rho(x^\mu)^C{}_D 
 \rho(x^\nu)^E{}_F u^Du^F
 \,,
\end{aligned}
\end{equation}
and an explicit calculation reveals that
\begin{equation}\label{InducedMetricGeneral}
\dd s^2  = u^4 u^0 \, \theta^R_0 \theta^R_0   +  u^4 u^\mu \, \theta^R_0 \theta^R_\mu - (u^4)^2 g^{\mu\nu} \theta^R_\mu \theta^R_\nu \,.
\end{equation}
$u^A$ cannot have a zero $4$ component, otherwise it would be left invariant by $G^*$. So, any choice of $u^A$ gives a  line element $ds^2$ of the general form~(\ref{General_R-invariant_metric}), which we studied in  detail in the previous section. The particular choice $u^A = \delta^A{}_4$ makes the matrix $g^{\mu\nu}$ used here and the one of Equation~(\ref{General_R-invariant_metric}) identical, otherwise, for other choices of fiducial vector, they are linearly related.

The matrix~(\ref{5DMatrix_hat_g}) is such that the submatrix $g^{\mu\nu}$ can be diagonalized with a linear redefinition of the first four indices, and the result is a matrix of the form:
\begin{equation}\label{5DMatrix_hat_g_v}
\hat{g}^{AB} = 
\left(
\begin{array}{ccccc}
\lambda^0 & 0 & 0 & 0 & v^0 
\\
0 & \lambda^1 & 0 & 0 & v^1
\\
0&0& \lambda^2 & 0 & v^2
\\
0& 0 & 0 & \lambda^3 & v^3
\\
v^0 & v^1 & v^2 & v^3 & 0
\end{array}
\right) \,,
\end{equation}
where $\lambda^\mu$ are the eigenvalues of  $g^{\mu\nu}$, and $v^\mu$ is the vector obtained by applying the similarity that diagonalizes  $g^{\mu\nu}$  to the vector $(1,0,0,0)$. However, now the transformed generators $\hat M'^{4B}$ do not provide a representation of the original ``timpelike'' $\kappa$-Minkowski algebra~(\ref{kappa-Minkowski_timelike}), but rather of an algebra isomorphic to it, in which the role of the special $x^0$ coordinate is played by a linear combination of the $x^\mu$ coordinates. The commutation relations are those of the \emph{generalized} $\kappa$-Minkowski algebra:
\begin{equation}\label{kappa-Minkowski_generalized}
[x^\mu,x^\nu] = \ii \left( v^\mu \, x^\nu - v^\nu \, x^\mu \right) \,.
\end{equation}
We therefore proved that the representation~(\ref{5DMatrix_hat_g}) of $\kappa$-Minkowski, when the matrix $g^{\mu\nu}$ is not chosen diagonal, is equivalent to a representation of the generalized $\kappa$-Minkowski algebra~(\ref{kappa-Minkowski_generalized}) with diagonal invariant matrix. This clarifies the meaning of the four parameters $v^\mu$ that appear in~(\ref{kappa-Minkowski_generalized}). In fact, the algebra~(\ref{kappa-Minkowski_generalized}) is  isomorphic to the standard $\kappa$-Minkowski algebra~(\ref{kappa-Minkowski_timelike}): it is sufficient to make the linear redefinition  $x^i \to v^0 x^i - v^i x^0$, $x^0 \to v_i x^i + \frac{1- \| \vec v \|^2}{v^0} x^0$ to show it. This does not, however, mean that all the noncommutative spaces corresponding to different choices of $v^\mu$ are completely equivalent. One has, in fact, to consider the whole group of symmetry that leaves the relations ~(\ref{kappa-Minkowski_generalized}) invariant under an inhomogeneous transformation of the form $x'^\mu = \Lambda^\mu {}_\nu x^\nu + a^\mu$, where $\Lambda^\mu{}_\nu$ and $a^\mu$ commute with $x^\mu$.
Any algebra of the following form:
\begin{equation}\label{CommRelGeneralizedKPgroup}
\begin{aligned}
[ a^\mu , a^\nu ] &= \ii \left( v^\mu \, a^\nu - v^\nu \, a^\mu \right) \,,
\\
[\Lambda ^\mu{}_\nu , \Lambda ^\rho{}_\sigma ] &= 0 \,,
\\
[\Lambda ^\mu{}_\nu , a^\gamma] &= \ii\left[\left( \Lambda^\mu_{\phantom{\mu}\alpha} \, v^\alpha - v^\mu \right)\Lambda^\gamma_{\phantom{\gamma}\nu}+\left( \Lambda^\alpha{}_\nu  \tilde{g}_{\alpha\beta}  -  \tilde{g}_{\nu\beta}  \right)v^\beta g^{\mu\gamma}\right] \,,
\end{aligned}
\end{equation}
(where we make no assumptions regarding the matrices $g$ and $\tilde{g}$ except that they are symmetric) will leave the commutation relations~(\ref{kappa-Minkowski_generalized}) invariant. However, the Jacobi rules are not  necessarily satisfied by these relations. In fact:
\begin{equation}
\begin{gathered}
[[\Lambda^\mu{}_\nu , a^\rho ] a^\sigma]+ 
[a^\rho , a^\sigma]\Lambda^\mu{}_\nu ]+
[[a^\sigma , \Lambda^\mu{}_\nu ] ,  a^\rho]
\\
=
 \left[ g^{\mu[\sigma} \Delta v^{\rho]}  \delta^\delta{}_\nu + g^{\delta[\rho} \Delta v^{\sigma]} \Lambda ^\mu{}_\nu + \Delta v^\mu g^{\delta[\sigma}  \Lambda ^{\rho]}{}_\nu + \Delta v^\delta g^{\mu[\sigma} \Lambda ^{\rho]}{}_\nu\right]  \tilde g_{\lambda \delta}v^\lambda ,
\end{gathered}\label{Jacobi}
\end{equation}
where $ \Delta v^\mu = v^\mu - g^{\mu \nu}  \tilde g_{\nu\rho} v^\rho$ and  $A^{[\mu\nu]} = A^{\mu\nu} - A^{\nu\mu}$. 
 A violation of the Jacobi rules would indicate a failure of associativity, which we would not know how to handle. We must therefore require~\eqref{Jacobi} to vanish in all of our systems.
An important case is when $g^{\mu\nu}$ is invertible and $\tilde g_{\mu\nu}$ is its inverse. Then the above expression is identically zero, because all the terms $ \Delta v^\mu = 0 $, \emph{for any choice of } $g^{\mu\nu}$.

We then understand better the meaning of our freedom to choose the parameters $v^\mu$: they select one linear combination of the noncommutative coordinates that plays a special role in the algebra, and we cannot simply re-label this direction ``$x^0$'' and reduce to the ``timelike'' case~(\ref{kappa-Minkowski_timelike}), because this linear transformation would change the matrix $g^{\mu\nu}$ that is left invariant by the isometries of our noncommutative spacetime. For example, depending on the initial choice of  $g^{\mu\nu}$, the redefinition that puts the algebra in the form~(\ref{kappa-Minkowski_timelike}) might end up making $x'^0$ into a spacelike coordinate. The simultaneous choice of $v^{\mu}$ and $g^{\mu\nu}$ determines a noncommutative geometry with a particular set of symmetries, and these fall into families that can be classified by looking at representations of the form~(\ref{5DMatrix_hat_g_v}), with a diagonal $g^{\mu\nu}$ and a generic $v^\mu$.\footnote{Alternatively, one could say that a model is specified by a choice of parameters $v^\mu$ and of matrix $\eta_{\mu\nu}$ in Eqs.~\ref{CommRelGeneralizedKPgroup}, but there are equivalence classes of \emph{physically equivalent} models. For instance, if two models are characterized by different $(v^\mu,g_{\mu\nu})$ and $(v'^\mu,g'_{\mu\nu})$, but there exists a linear redefinition of the generators $x^\mu$ that transforms $v^\mu$ into $v'^\mu$, and, at the same time, $g^{\mu\nu})$ into $g'^{\mu\nu}$, then the two models are the same, we are just labeling its coordinates in two different ways.} Each choice of the four eigenvalues $\lambda^\mu$ gives a 4-parameter family (dependent on $v^\mu$) of inequivalent models. We can then list these inequivalent models and describe the geometry of the corresponding momentum spaces.

\section{Nondegenerate geometries of momentum space}
\label{SectionMomentumSpaceGeometries}

We are ready to list the possible \emph{nondegenerate} geometries of momentum spaces one can build for the $\kappa$-Minkowski noncommutative spacetime. There are also many degenerate cases, in which the momentum space manifold is not \emph{topologically} 4-dimensional, but we prefer to focus on the nondegenerate ones, which have a chance to be useful in model building. Although all the cases we list have topological dimension four, their geometry might be degenerate, in the sense that the induced metric might have one or more zero eigenvalues.

Let us summarize, for clarity, the algorithm we use to write different representations of the $\kappa$-Minkowksi coordinate algebra. Given an ambient metric of the form (\ref{5DMatrix_hat_g_v}), we can write a matrix representation of the 10-dimensional Lie algebra that preserves this matrix, in the sense of satisfying commutation relations of the form~(\ref{5DOrthogonalAlgebra}), as
\begin{equation}
(\hat M^{AB})^a{}_b = i \left( \hat g^{A a} \delta^B{}_b - \hat g^{B a} \delta^A{}_b \right)\,,
\end{equation}
and then the components $\rho(x^\mu) = \hat M^{4 \mu}$ close a generalized $\kappa$-Minkowski algebra of the form~(\ref{kappa-Minkowski_generalized}). Explicitly,
\begin{equation}\label{Final_kappa_Minkowski_representation}
\begin{aligned}
\rho(x^0) =
\ii \left(
\begin{array}{ccccc}
v^0 & 0 & 0 & 0 & -\lambda^0
\\
v^1 & 0 & 0 & 0 & 0
\\
v^2 & 0 & 0 & 0 & 0
\\
v^3 & 0 & 0 & 0 & 0
\\
0 & 0 & 0 & 0 & -v^0
\end{array}
\right) 
\,,
\qquad
\rho(x^1) =
\ii \left(
\begin{array}{ccccc}
0 & v^0 & 0 & 0 & 0
\\
0 & v^1 & 0 & 0 & - \lambda^1
\\
0 & v^2 & 0 & 0 & 0
\\
0 & v^3 & 0 & 0 & 0
\\
0 & 0 & 0 & 0 & -v^1
\end{array}
\right) \,,
\\
\rho(x^0) =
\ii \left(
\begin{array}{ccccc}
0 & 0 & v^0 & 0 & 0
\\
0 & 0 & v^1 & 0 & 0
\\
0 & 0 & v^2 & 0 & - \lambda^2
\\
0 & 0 & v^3 & 0 & 0
\\
0 & 0 & 0 & 0 & -v^2
\end{array}
\right) 
\,,
\qquad
\rho(x^1) =
\ii \left(
\begin{array}{ccccc}
0 & 0 & 0 & v^0 & 0
\\
0 & 0 & 0 & v^1 & 0
\\
0 & 0 & 0 & v^2 & 0
\\
0 & 0 & 0 & v^3 & - \lambda^3
\\
0 & 0 & 0 & 0 & -v^3
\end{array}
\right) \,.
\end{aligned}
\end{equation}
This is the most generic representation of the generalized $\kappa$-Minkowski commutation relations, and it is easy to verify that it reduces to the previously-known cases for specific choices of $\lambda^\mu$ and $v^\nu$ (possibly after a permutation of the columns/rows). The values of  $\lambda^\mu$ determine the geometry of the associated momentum space.

The orbits $Y^A(p_\mu) = G^*(p_\mu)^A{}_B u^B$ of the elements $G^*(p_\mu) = \exp [ i p_i \rho(x^i)] \, \exp [ i p_0 \rho(x^0)]$ are our momentum spaces, and, as we showed above, it doesn't matter which point $u^A$ in the ambient $\mathbbm{R}^5$ we start from, we can always find a choice of $\lambda^\mu$ parameters for which the orbit includes the point $u^A = (0,0,0,0,1)$, so we might as well always use this as our starting point.

By construction, the embedding coordinates $Y^A$ satisfy the quadratic equation
\begin{equation}
\hat g_{AB} Y^A Y^B = \hat g_{AB} u^A u^B = \frac{\lambda^0\lambda^1\lambda^2\lambda^3}{\det \hat g^{AB} } = \text{\it const.} \,,
\end{equation}
and, moreover, the fifth coordinate satisfies
\begin{equation}
Y^4 = e^{p_\mu v^\mu} > 0 \,,
\end{equation}
which is the same as relation $X^0 + X^4 >0 $ from above, written in different coordinates. We can now list the inequivalent cases one gets with different choices of $\lambda^\mu$, and describe the geometry of the corresponding momentum spaces.

\newpage

\subsection*{Half of de Sitter momentum space}

\begin{wrapfigure}[10]{r}{0.35\textwidth}\begin{center}
\vspace{-48pt}
\includegraphics[width=0.35\textwidth]{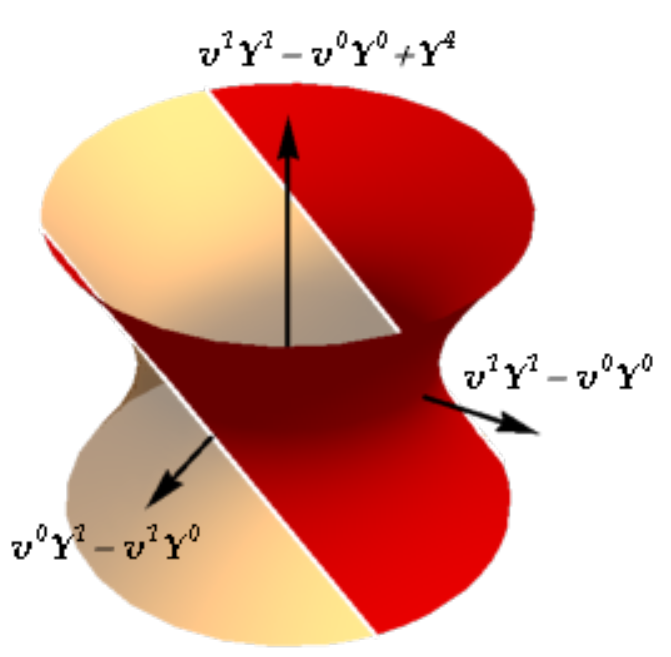}
\end{center}
\caption{\sl Half of de Sitter space covered by the $p^\mu$ coordinates in 1+1 dimensions (in red).}\label{dSFig}
\end{wrapfigure}

This case has already been sketched in the introduction, let us work it out again in our new setup. With the following choice of ambient metric: 
\begin{equation}
\hat{g}^{AB} = 
\left(
\begin{array}{ccccc}
-1 & 0 & 0 & 0 & v^0 
\\
0 & 1 & 0 & 0 & v^1
\\
0&0& 1 & 0 & v^2
\\
0& 0 &0& 1 & v^3
\\
v^0 & v^1 & v^2 & v^3 & 0
\end{array}
\right)
\end{equation}
the generators $\hat M^{4\mu}$ of~(\ref{5DOrthogonalAlgebra}) close the generalized $\kappa$-Minkowski algebra, and their exponentialization $G^*(p)^A{}_B$ acts on the fiducial vector $u^A = \delta^A{}_4$ in such a way that the embedded hypersurface it defines, $Y^A(p^\mu) = G^*(p)^A{}_B u^B$,  is the $Y^4 >0$ half of a de Sitter hyperboloid. The induced line element on the embedded hypersurface is
\begin{equation}
\dd s^2 = - \dd p_0^2 + e^{2 p_0/\kappa} (\dd p_1^2 + \dd p_2^2 + \dd p_3^2) \,,
\end{equation}
which is the de Sitter metric in flat slicing, well-known to be a half-cover of the hyperboloid~\cite{Pascu}. In Fig.~\ref{dSFig} is a graphic representation of the embedded manifold in the 1+1-dimensional case.

The homogeneous isometries of this momentum space are given by the Lorentz group $SO(3,1)$, to which the matrices $\Lambda^\mu{}_\nu$ of Eqs.(\ref{CommRelGeneralizedKPgroup}) belong to in this case, because they leave the Minkowski metric and its inverse invariant, \emph{i.e.} $\Lambda^\mu{}_\rho \Lambda^\nu{}_\sigma \eta_{\mu\nu} = \eta_{\rho\sigma}$ and $\Lambda^\mu{}_\rho \Lambda^\nu{}_\sigma \eta^{\rho\sigma} = \eta^{\mu\nu}$. Therefore, in this case, we are dealing with a quantum group of isometries of the  noncommutative spacetime~(\ref{kappa-Minkowski_generalized}), of  the form~(\ref{CommRelGeneralizedKPgroup}) with $\tilde g_{\mu\nu} = \text{diag}(-1,1,1,1) = \tilde g^{\mu\nu}$. This is a quantum deformation of the Poincar\'e group $ISO(3,1)$.

\subsection*{Half of anti-de Sitter space}

\begin{wrapfigure}[10]{r}{0.35\textwidth}\begin{center}
\vspace{-48pt}
\includegraphics[width=0.35\textwidth]{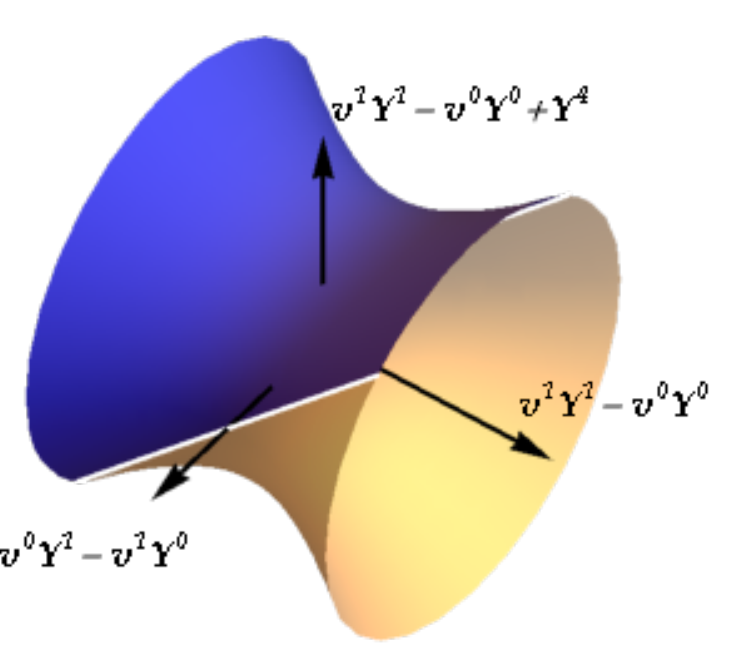}
\end{center}
\caption{\sl Half of anti-de Sitter space covered by the $p^\mu$ coordinates in 1+1 dimensions (in blue).}
\end{wrapfigure}
By placing the minus in one of the diagonal components corresponding to the $1$, $2$ or $3$ axes, for example:
\begin{equation}
\hat{g}^{AB} = 
\left(
\begin{array}{ccccc}
1& 0 & 0 & 0 & v^0 
\\
0 &-1 & 0 & 0 & v^1
\\
0&0& 1 & 0 & v^2
\\
0& 0 &0& 1 & v^3
\\
v^0 & v^1 & v^2 & v^3 & 0
\end{array}
\right)
\end{equation}
we get another Lorentzian momentum space manifold, whose induced metric has negative constant curvature:
\begin{equation}
\dd s^2 =  \dd p_0^2 + e^{2 k_0/\kappa} (- \dd p_1^2 + \dd p_2^2 + \dd p_3^2) \,,
\end{equation}
(the minus could be in front of any of the $\dd p_1^2$, $\dd p_2^2$, or $\dd p_3^2$ terms).
Again, the sum $Y^4$ is positive for all values of $p^\mu$, so we are dealing with half a hyperboloid. Just like in the case above, the region covered by these coordinates is not Lorentz-invariant, and one has to assume a nontrivial global topology (like the elliptic topology of de Sitter) in order to recover Lorentz invariance~\cite{Mercati:2018ruw,Mercati:2018fba}.

The local homogenous isometry group is again $SO(3,1)$. The quantum group of symmetries is then a deformation of  $ISO(3,1)$, but with the timelike direction being along the axis $1$ (or $2$, or $3$, according to the choice of $\hat g^{AB}$). In other words, this model is identical to the one of the previous point, except that the Lorentz matrices satisfy $\Lambda^\mu{}_\rho \Lambda^\nu{}_\sigma \tilde g_{\mu\nu} = \tilde g_{\rho\sigma}$ and $\Lambda^\mu{}_\rho \Lambda^\nu{}_\sigma \tilde g^{\rho\sigma} = \tilde g^{\mu\nu}$ , where $\tilde g_{\mu\nu} = \text{diag}(1,-1,1,1) = \tilde g^{\mu\nu}$, \emph{i.e.}, it has a negative $11$, $22$ or $33$ component, as opposed to the $00$ component.

\subsection*{Riemannian hyperbolic momentum space} 

\begin{wrapfigure}[12]{r}{0.35\textwidth}\begin{center}
\vspace{-48pt}
\includegraphics[width=0.35\textwidth]{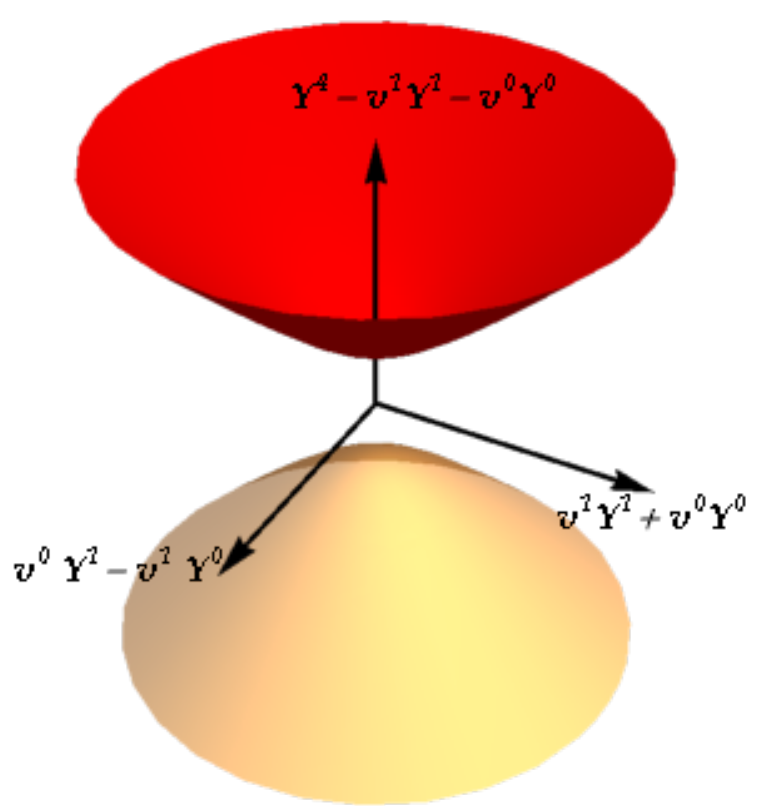}
\end{center}
\caption{\sl Riemannian hyperbolic momentum space in 1+1 dimensions. The $p^\mu$ coordinates cover only the red sheet.}
\end{wrapfigure}
With a Euclidean signature for the 4D submatrix:
\begin{equation}
\hat{g}^{AB} = 
\left(
\begin{array}{ccccc}
1 & 0 & 0 & 0 & v^0 
\\
0 & 1 & 0 & 0 & v^1
\\
0&0& 1 & 0 & v^2
\\
0& 0 &0& 1 & v^3
\\
v^0 & v^1 & v^2 & v^3 & 0
\end{array}
\right)
\end{equation}
we get a Euclidean-signature hyperboloid for momentum space. The condition $Y^4 >0$ still holds, but that includes the entirety of one of the two disconnected sheets of the hyperboloid. The attractive feature of this momentum space is that the condition $Y^4 >0$ does not identify a region with boundary, and, what is more, the region it identifies is invariant under `Lorentz' transformations, unlike what happens in the two cases listed above. We have a momentum space that is closed under 4-dimensional rotations, which might represent an appealing simplicity feature for the study of Euclidean QFT.

The induced metric on this hyperboloid is:
\begin{equation}
\dd s^2 = \dd p_0^2 + e^{2 k_0/\kappa} (\dd p_1^2 + \dd p_2^2 + \dd p_3^2)\,.
\end{equation}

The homogeneous isometries are $SO(4)$ rotations, and the quantum group associated to this momentum space is~(\ref{CommRelGeneralizedKPgroup}) with $\tilde g_{\mu\nu} = \text{diag}(1,1,1,1) = \tilde g^{\mu\nu}$, which is a deformation of $ISO(4)$, the group of 4-dimensional Euclidean transformations.

\subsection*{Two-time hyperbolic space}

With a 4D sub-matrix with two positive and two negative eigenvalues:
\begin{equation}
\hat{g}^{AB} = 
\left(
\begin{array}{ccccc}
-1 & 0 & 0 & 0 & v^0 
\\
0 & -1 & 0 & 0 & v^1
\\
0&0& 1 & 0 & v^2
\\
0& 0 &0& 1 & v^3
\\
v^0 & v^1 & v^2 & v^3 & 0
\end{array}
\right)
\end{equation}
we get a manifold with two temporal and two spatial directions, with metric:
\begin{equation}
\dd s^2 = -\dd p_0^2 + e^{2 p_0/\kappa} (-\dd p_1^2 + \dd p_2^2 + \dd p_3^2) \,,
\end{equation}
(again, we have three distinct cases, related by exchanges of the $1$, $2$, $3$ axes).  Whatever the values of $v^\mu$, we can always rotate our axes $Y^i \to Y'^i$  so that this manifold corresponds to region $Y^4 >0$ of the quadric
\begin{equation}
(Y^4)^2 + 2 Y^0 Y^4 - (Y'^1)^2  +  (Y'^2)^2 + (Y'^3)^2 =  1 \,.
\end{equation}
This region has a finite boundary at, $Y^4=0$ in correspondence of the submanifold $(Y'^2)^2 + (Y'^3)^2 =  1 + (Y'^1)^2 $.

It does not make sense to try and represent this momentum space in $1+1$ dimensions as a submanifold of $\mathbbm{R}^3$, so we won't show, in this case, a pictorial representation.

The homogeneous isometries of this manifold close an $SO(2,2)$ group. Therefore this momentum space is associated to a noncommutative spacetime which is symmetric under a quantum deformation of  $ISO(2,2)$ (\emph{i.e.}~(\ref{CommRelGeneralizedKPgroup}) with $\tilde g_{\mu\nu} = \text{diag}(-1,-1,1,1) = \tilde g^{\mu\nu}$).

\subsection*{Light cone momentum spaces}

\begin{wrapfigure}[13]{r}{0.35\textwidth}\begin{center}
\vspace{-48pt}
\includegraphics[width=0.35\textwidth]{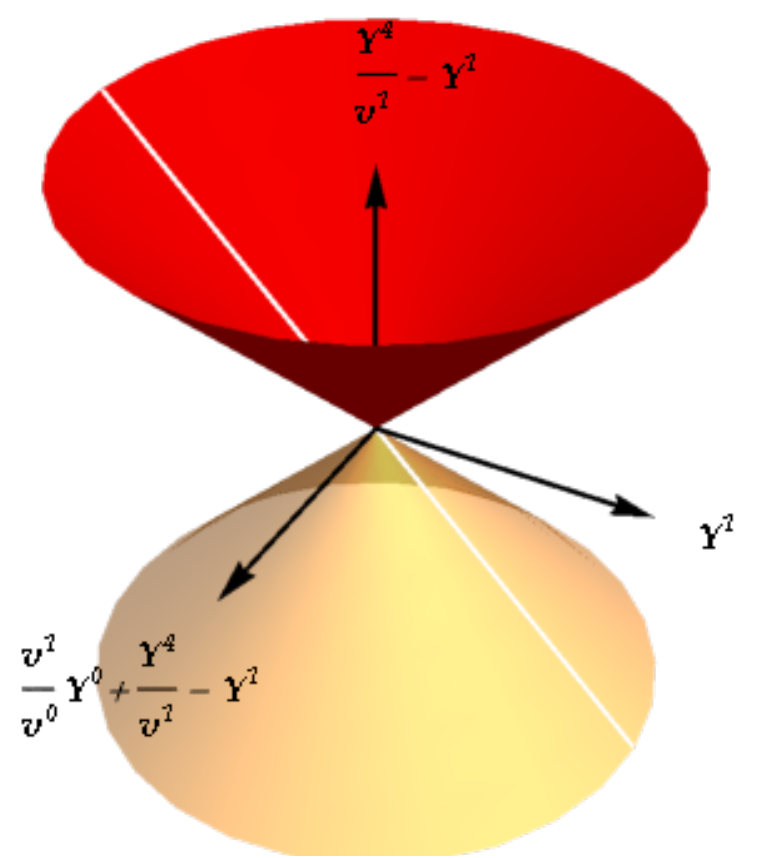}
\end{center}
\caption{\sl Light-cone momentum space. The $p^\mu$ coordinates cover only the red sheet, with the exclusion of the white line.} \label{ConeFig}
\end{wrapfigure}
We get an interesting momentum space if we set $\lambda^0 =0$. The other $\lambda^\mu$'s can have either Lorentzian or Euclidean signature:
\begin{equation}
\hat{g}^{AB} = 
\left(
\begin{array}{ccccc}
0 & 0 & 0 & 0 & v^0 
\\
0 & \pm 1 & 0 & 0 & v^1
\\
0&0& 1 & 0 & v^2
\\
0& 0 &0& 1 & v^3
\\
v^0 & v^1 & v^2 & v^3 & 0
\end{array}
\right) \,,
\end{equation}
(any of the $\lambda^1$, $\lambda^2$ or $\lambda^3$ could be negative, or two of them could be negative and one positive, without changing the result). In fact, the $\hat g_{AB}$-norm of $u^A$ is, in this case, zero, and we get a light cone in the embedding space. Suppressing the $3$ direction, we can see it represented in Fig.~(\ref{ConeFig}): the condition $Y^4>0$ selects only the ``future'' half of the cone, and it further cuts a line off of it (see Fig.~\ref{ConeFig}). Notice how this makes it topologically equivalent to all the other momentum spaces we encountered so far (although the metric of the $AN(3)$ group manifold is not unique, the topology is uniquely determined by its Lie group structure).

The induced metric is degenerate:
\begin{equation}\label{carroll_example_metric}
\dd s^2 = e^{2 p_0/\kappa} (\pm \dd p_1^2 + \dd p_2^2 + \dd p_3^2) \,,
\end{equation}
and its homogeneous isometries are a contraction of the Lorentz group $SO(3,1)$ (or $SO(2,2)$ in the case with the minus sign) into $ISO(3)$ (or $ISO(2,1)$). This is the contraction in which the speed of light is sent to zero,
and the corresponding inhomogeneous group is the Carroll group~\cite{LLeblondCarroll,Duval:2014uoa} $\text{Carr}(3,1)$ (or, in the case with the minus sign, a version of the Carroll group in which the spatial directions are Lorentzian). This quantum group is  still of the form~(\ref{CommRelGeneralizedKPgroup}), but with $\tilde g^{\mu\nu} = \text{diag}(0,1,1,1)$ and $\tilde g_{\mu\nu}= \text{diag}(1,0,0,0) $ (this is a possible way to define the Carroll group, as the group of inhomogeneous transformations $x^\mu \to \Lambda^\mu{}_\nu x^\nu + a^\mu$ where the matrices $\Lambda^\mu{}_\nu$ preserve a degenerate metric with one zero eigenvalue, and their transposes preserve a `complementary' metric with three zero eigenvalues. The Galilei group can be obtained similarly, by exchanging the roles of the two metrics).  In this case, since the metric is not invertible, the Jacobi relation~\eqref{Jacobi} is not automatically satisfied. It turns out that for the Carroll case it holds when the vector $v$ is time-like, while the Galilei case requires it to be space-like.

We then found the momentum space associated to a noncommutative spacetime with the commutation relations~(\ref{kappa-Minkowski_generalized}) of $\kappa$-Minkowski (generalized), and a deformation of the Carroll group as quantum group of symmetry. Such a deformation has been already considered recently in~\cite{Daskiewicz_Carroll_Galilei}, and specifically in the $\kappa$-deformed case in~\cite{kappaGalileiCarroll}.
We can add, to the discussion of~\cite{kappaGalileiCarroll}, the observation that the momentum space associated to this Carrollian $\kappa$-deformed noncommutative space is (part of) a cone.

\subsection*{Half Minkowski momentum space}
 
As shown in Sec.~(\ref{Right-invariant-metrics-sec}), one can have a flat momentum space with a nondegenerate metric by choosing a $\hat{g}^{AB}$ matrix of the form:
\begin{equation}
\hat{g}^{AB} = 
\left(
\begin{array}{ccccc}
0 & 1 & 0 & 0 & v^0 
\\
1 & 0 & 0 & 0 & v^1
\\
0&0& 1 & 0 & v^2
\\
0& 0 &0& 1 & v^3
\\
v^0 & v^1 & v^2 & v^3 & 0
\end{array}
\right) \,,
\end{equation}
modulo exchanges of $1$, $2$ $3$ axes.
The induced metric in this case is:
\begin{equation}
\label{FlatMomentumSpaceMetric}
\dd s^2 = 2 \, e^{2 p_0/\kappa} \dd p_0 \dd p_1 + e^{2 p_0/\kappa} (\dd p_2^2 + \dd p_3^2) \,,
\end{equation}
which is locally diffeomorphic to the Minkowski metric. However, the global geometry can't be that of Minkowski. Drawing an embedding diagram of the group orbits in this case won't help much, so we resort to studying the metric in the $p_\mu$ coordinates. Begin with the Minkowski metric:
\begin{equation}\label{MinkowskiMetric}
\dd s^2 = - \dd t^2 + \dd x^2+ \dd y^2+ \dd z^2\,,
\end{equation}
and make the following coordinate transformation, consisting of a special conformal transformation followed by a translation:
\begin{equation}
\begin{aligned}
 t &=-\frac{\sqrt{2} \left(-\left(a+\frac{1}{2}\right)^2+b^2+p_2^2+p_3^2+a+\frac{1}{2}\right)}{a+b} \,, 
 \\
 x &=-\frac{\sqrt{2} \left(\left(a+\frac{1}{2}\right)^2-b^2-p_2^2-p_3^2+b\right)}{a+b}  \,, 
 \\
 y &=-\frac{\sqrt{2} p_2}{a+b} \,, 
  \\
 z &=-\frac{\sqrt{2} p_3}{a+b} 
\end{aligned}
\end{equation}
where
\begin{equation}
 a =\frac{\tau - p_1}{\sqrt{2}} \,,  \qquad b=\frac{\tau+p_1 }{\sqrt{2}} \,.
\end{equation}
Then the metric~(\ref{MinkowskiMetric})takes the form
\begin{equation}
\dd s^2 = \frac 1 {\tau^2} \left( 2 \dd \tau \dd p_1 + \dd p_2^2 + \dd p_3^2 \right) \,,
\end{equation}
which can be put in the form~(\ref{FlatMomentumSpaceMetric}) with the redefinition $\tau = e^{- p_0 / \kappa}$. Now, to find the boundary of our coordinate patch, it is sufficient to observe that $-(t+x) = \frac{e^{p_0/\kappa} }{2}+\sqrt{2} > \sqrt{2}$: we are dealing with half of Minkowski space.

We found a third model whose quantum group is a deformation of Poincar\'e. The difference between the first two, which had, respectively, a de Sitter and an anti-de Sitter momentum space, was that, among the noncommutative coordinates $x^\mu$, the one corresponding to the time-like direction was $x^0$ in the first case, and one of the $x^i$ in the second case. This third case is at the interface between the two previous ones: the timelike coordinate is the difference between $x^0$ and one of the $x^i$, while their sum is spacelike.

Notice that this has nothing to do with the form of the commutation relations~(\ref{kappa-Minkowski_generalized}), where the parameters $v^\mu$ are completely free and determine which linear combination of $x^\mu$'s act as a dilatation for the other ones (which close an Abelian subalgebra). One can have a model with, say, anti-de Sitter momentum space and timelike $x^1$ coordinate, with commutation relations that make $x^0$ the dilatation-like generator.

\subsection*{Singular cases}

There are cases which correspond to  quantum deformations of sensible Lie groups, in which the matrix $\hat g^{AB}$ is singular. In particular, if two or more diagonal elements $\lambda^\mu$ are zero, the matrix is singular, and one cannot write the momentum space manifold as a quadratic form in the ambient space. We can however study it as an embedded hypersurface, by exponentializing the representation~(\ref{Final_kappa_Minkowski_representation}). In the case $\lambda^0=1$, $\lambda^i=0$, we get an embedded manifold of the form
\begin{equation}
Y^A = \left( \frac{u-\frac{u^2}{2 w}}{v^0},-\frac{u^2 v^1}{\left(v^0\right)^2 (2 w)},-\frac{u^2 v^2}{\left(v^0\right)^2 (2 w)},-\frac{u^2 v^3}{\left(v^0\right)^2 (2 w)},w  \right) 
\end{equation}
where $u = e^{k_0  v^0} -1$ and $w = e^{k_\mu v^\mu}$, which is clearly two-dimensional. This choice of parameters $\lambda^\mu$ is dual to that of the Carroll group and therefore corresponds to a quantum deformation of the Galileian group. We find that the corresponding momentum space is topologically degenerate, in the sense that it is not four-dimensional. The same happens if we choose one of the $\lambda^i $ to be the only nonzero one (which corresponds to a sort of Galilei group in which the spatial rotations and translations are replaced by a Lorentzian $ISO(2,1)$ subgroup).

The cases with two zero and two nonzero $\lambda^\mu$'s are all degenerate in the same sense, as can be straightforwardly verified by considering the $5 \times 4$ matrix $\partial_{k_\mu} Y^A$, which has no $4 \times 4$ submatrix with nonzero determinant.

\section{Conclusions and Outlook}

We found that the $\kappa$-Minkowski noncommutative spacetime, or rather its four-parameter generalization~(\ref{kappa-Minkowski_generalized}), admits a plethora of nondegenerate momentum spaces, all with the same topology but with different metric properties and different signatures (including degenerate ones). These momentum spaces are each associated to a different
quantum group describing the symmetries of the same (family of) noncommutative spacetimes~(\ref{kappa-Minkowski_generalized}). These groups are characterized by the relations~(\ref{CommRelGeneralizedKPgroup}), where $\tilde g_{\mu\nu}$ and  $\tilde g^{\mu\nu}$  are numerical matrices which, when invertible, are one the inverse of the other. In these invertible cases one has a quantum deformation of the  Poincar\'e group $ISO(3,1)$, the Euclidean group $ISO(4)$, or the group $ISO(2,2)$. In the first case (Poincar\'e), we can associate three different momentum spaces to the model, with very different geometries: a half de Sitter, half anti-de Sitter or  a half Minkowski space. These three cases are all associated with a Poincar\'e-invariant $\kappa$-Minkowski noncommutative spacetime, but they are distinguished by which of the $x^\mu$ coordinate (or rather, which of the four conjugate momenta) corresponds to the timelike direction in momentum space. If it is $x^0$ (or a Lorentz transformation thereof), we have a de Sitter momentum space, when it is connected by a Lorentz transformation to one of the $x^1$, $x^2$, $x^3$ coordinates, we have anti-de Sitter, an if it is a $45^\circ$ combination of $x^0$ with one of the $x^i$, we have a Minkowski space. Notice that this has nothing to do with the choice of parameters $v^\mu$, which determine which linear combination of the $x^\mu$ acts as a dilatation upon the other ones, which close an Abelian subalgebra. One can have any of the three momentum spaces with any choice of $v^\mu$. It is also worth mentioning how the light-cone case represents somehow an interface between the dS and AdS cases: these two momentum spaces have, respectively, positive and negative curvature, while the light cone has zero curvature. Similarly, the timelike combination of coordinates $x^\mu$ corresponding to the light-cone momentum space is the limit case between  linear combinations of the form $x^0 \cosh \tau + x^i \sinh \tau $ for some (possibly $SO(3)$-rotated) $i$ and a real value of the parameter $\tau$, which all admit a de Sitter momentum space, and  combinations of the form   $x^0 \sinh \tau + x^i \cosh \tau $, which correspond to an anti-de Sitter momentum space. It would be interesting to consider $\kappa$-deformed models in regimes whose spacetime symmetries are described by the Carroll group, \emph{i.e.} in presence of strong gravity (for an exploration of the phenomenology of such models see, for example,~\cite{Gravityquantumspacetime}).

Furthermore, we found a previously-unkown momentum space with the geometry of the (forward) light-cone of a $5D$ Minkowski space (with one line removed). Strikingly, this momentum space corresponds to a quantum group of symmetries based on the Carroll group, in which the matrices $\tilde g^{\mu\nu}$ and  $\tilde g_{\mu\nu}$ left invariant by the homogeneous transformations $\Lambda^\mu{}_\nu$ have, respectively, one and three zero eigenvalues (in orthogonal directions). There is also an associated case in which the momentum space is still a cone, but the ambient space has signature $(2,2)$. This corresponds to a Carroll-like group in which the matrix $\tilde g_{\mu\nu}$  has one zero, one negative and two positive eigenvalues. We stress that  momentum spaces for $\kappa-$Minkowski with a (quantum) Caroll group of symmetry have never been studied before.

Finally, we explored the natural question: if we have a momentum space associated to the $\kappa$-Carroll group, don't  we have one associated to a $\kappa$-Galileian group, which can be defined through relations~(\ref{CommRelGeneralizedKPgroup}) with the roles of $\tilde g^{\mu\nu}$ and  $\tilde g_{\mu\nu}$ exchanged? The answer appears to be no, because the corresponding momentum space manifold is not four-dimensional (it is degenerate in a topological sense, not only metrically). The same holds for a group in which $\tilde g^{\mu\nu}$ and  $\tilde g_{\mu\nu}$ have both two zero and two nonzero eigenvalues.

This, we believe, exhausts the study of all the interesting momentum spaces that can be constructed on $\kappa$-Minkowski, and their associated quantum groups of space(time) symmetries.
Our analysis also unifies past approaches to $\kappa$-Momentum spaces based on right-invariant metrics~\cite{DanijelJurman2017} and matrix representations of $\kappa$-Minkowski~\cite{KowalskiGlikman:2003we,Arzano:2014jfa,Mercati:2018hlc,Ballesteros:2017pdw}, clarifying that there is a right-invariant metric on the $AN(3)$ group for each one of the momentum spaces that we identified, and an associated $5D$ matrix representation of the algebra~(\ref{kappa-Minkowski_generalized}) in terms of a subalgebra of generators of a Lie group of homogeneous isometries of an ambient $5D$ metric. The linear complement of this subalgebra always closes a subalgebra which coincides with the (Lie algebra of the) homogeneous part of the quantum group of symmetries of $\kappa$-Minkowski associated to the chosen momentum space.

\subsubsection*{Ackowledgments}

We thank J. Kowalski-Glikman for pointing out Ref.~\cite{Blaut2004} and the results therein, which partly superposed with the previous version of our paper, and stimulated us to develop the new approach described in the present version.
FL and MM acknowledge support from the INFN Iniziativa Specifica GeoSymQFT. FM the Action CA18108 QG-MM from the European Cooperation in Science and Technology (COST) and partial support from the Foundational Questions Institute (FQXi). FL the Spanish MINECO under project MDM-2014-0369 of ICCUB (Unidad de Excelencia ‘Maria de Maeztu’), grant FPA2016-76005-C2-1-P.

\providecommand{\href}[2]{#2}\begingroup\raggedright\endgroup

\end{document}